\documentclass[a4paper,11pt]{article}
\usepackage{pos}

\title{Deep-Inelastic Scattering with LHC Neutrinos}

\author*[a,b]{Juan Rojo}

\affiliation[a]{Nikhef Theory Group, Science Park 105, 1098 XG Amsterdam, The Netherlands}
\affiliation[v]{Department of Physics and Astronomy, Vrije Universiteit, NL-1081 HV Amsterdam, The Netherlands}

\emailAdd{j.rojo@vu.nl}

\abstract{The observation of neutrinos produced in LHC collisions by the far-forward FASER and SND@LHC experiments in 2023 herald the new era of collider neutrino physics. 
These high-intensity forward neutrino fluxes from proton-proton LHC collisions enable novel opportunities in QCD, neutrino physics, BSM searches, and astroparticle physics.
In this contribution, we briefly review the physics potential of neutrino deep-inelastic scattering (DIS) measurements at the LHC to shed light on proton and nuclear structure, both for ongoing far-forward experiments and for the proposed Forward Physics Facility (FPF) to operate concurrently with the HL-LHC.
We consider the reach of LHC neutrino DIS to scrutinize the mechanisms of forward hadron production and constrain QCD in the ultra-small-$x$ region.
We also discuss how neutrino DIS measurements at these far-forward detectors enhance direct and indirect BSM searches at the HL-LHC.}

\FullConference{31st International Workshop on Deep Inelastic Scattering (DIS2024)\\
 8–12 April 2024\\
Grenoble, France\\}

\newcommand{\be}{\begin{equation}}
\newcommand{\ee}{\end{equation}}
\newcommand{\bea}{\begin{eqnarray}}
\newcommand{\eea}{\end{eqnarray}}
\newcommand{\bi}{\begin{itemize}}
\newcommand{\ei}{\end{itemize}}
\newcommand{\ben}{\begin{enumerate}}
\newcommand{\een}{\end{enumerate}}

\def\frac#1#2{{{#1}\over {#2}}}
\def\gsim{\mathrel{\rlap{\lower4pt\hbox{\hskip1pt$\sim$}}
    \raise1pt\hbox{$>$}}}       
\def\lsim{\mathrel{\rlap{\lower4pt\hbox{\hskip1pt$\sim$}}
    \raise1pt\hbox{$<$}}}

\begin{document}

\maketitle

\paragraph{Introduction.}
The discovery of LHC neutrinos in 2023 by the far-forward experiments FASER~\cite{FASER:2023zcr} and SND@LHC~\cite{SNDLHC:2023pun} herald the beginning of the collider neutrino era in particle physics.
Shortly thereafter, FASER$\nu$~\cite{FASER:2024hoe} presented a first laboratory measurement of the neutrino interaction cross-section at TeV energies.
These groundbreaking results with LHC neutrinos are the harbinger of the many exciting opportunities becoming available in several topics in particle, hadronic, and astroparticle physics, including searches for new physics beyond the Standard Model (BSM), see~\cite{Anchordoqui:2021ghd,Feng:2022inv} and references therein. 
Both FASER and SND@LHC will take data for the rest of the ongoing LHC Run-3.
FASER has been approved to also operate (without the FASER$\nu$ component) during Run-4, while SND@LHC is considering novel locations in the LHC accelerator complex to increase its exposure to the neutrino fluxes.
Upgrades of these far-forward experiments, such as the proposed FASER($\nu$)2, as well as new ones, such as FLArE~\cite{Batell:2021blf}, may be installed in a new dedicated Forward Physics Facility (FPF)~\cite{Anchordoqui:2021ghd,Feng:2022inv} to operate concurrently with the HL-LHC.

\begin{figure}[h]
    \centering
\includegraphics[width=0.59\textwidth,angle=-90]{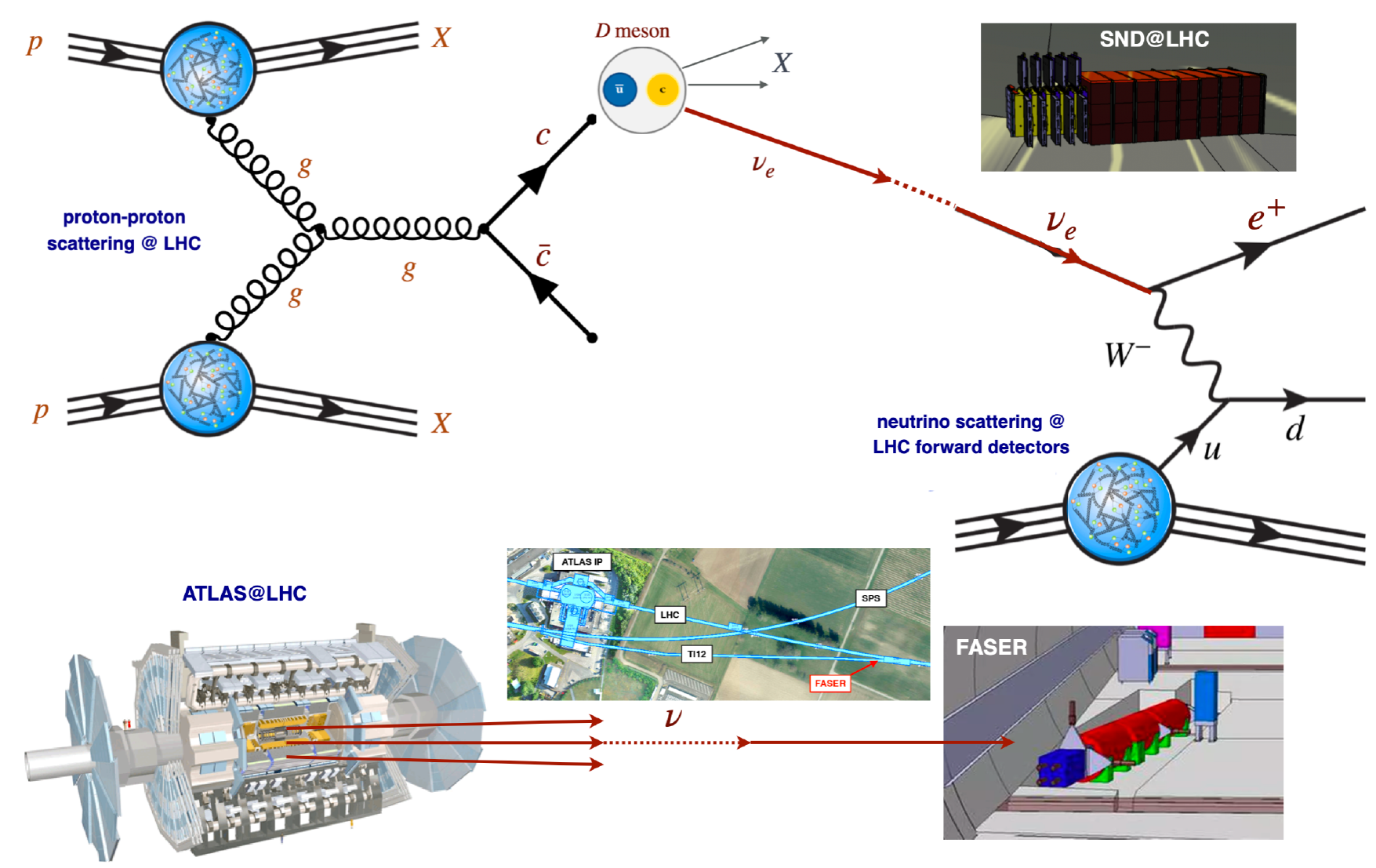}
    \caption{Schematic representation of forward neutrino production at the LHC and its subsequent deep-inelastic
    scattering at far-forward detectors.
    We display the mechanism dominating the high-energy $\nu_e$ fluxes,  namely $D$-meson production, which is sensitive to the gluon PDF down to $x\sim 10^{-7}$.
    }
    \label{fig:procs-fig1}
\end{figure}

Fig.~\ref{fig:procs-fig1} provides an schematic representation of forward neutrino production at the LHC and its subsequent deep-inelastic scattering at the far-forward detectors.
We display the mechanism dominating the high-energy $\nu_e$ fluxes,  namely $D$-meson production~\cite{Kling:2021gos,Buonocore:2023kna}, which is sensitive to the gluon PDF down to $x\sim 10^{-7}$ beyond the coverage of any other laboratory experiment. 
Measurements of neutrino event yields at the LHC far-forward experiments hence provide information both on the production mechanisms of forward neutrinos as well as on their interactions with the target material via the DIS process.
The resultant understanding of TeV-energies neutrino production and scattering enabled by the LHC far-forward experiments feeds directly into oscillation and astroparticle physics  studies at experiments such as IceCube/DeepCore~\cite{IceCube:2017lak,IceCube:2014stg} and KM3NET~\cite{KM3NeT:2021ozk,KM3Net:2016zxf}.

\paragraph{Constraints on proton and nuclear structure from neutrino DIS.}

The impact of ongoing and future LHC far-forward neutrino experiments on proton and nuclear structure via the DIS process has been quantified in~\cite{Cruz-Martinez:2023sdv}.
For this study, we generated dedicated projections for DIS structure function measurements at these experiments (similarly to the HL-LHC PDF projections of~\cite{AbdulKhalek:2018rok}), finding that up to one million electron-neutrino and muon-neutrino DIS interactions could be recorded by the end of the LHC operations.
The kinematic region in $(x,Q^2)$ of these neutrino collisions overlaps with that of the Electron-Ion Collider~\cite{AbdulKhalek:2021gbh}, while accessing complementary flavour combinations.

\begin{figure}[h]
    \centering
\includegraphics[width=0.47\textwidth,angle=-90]{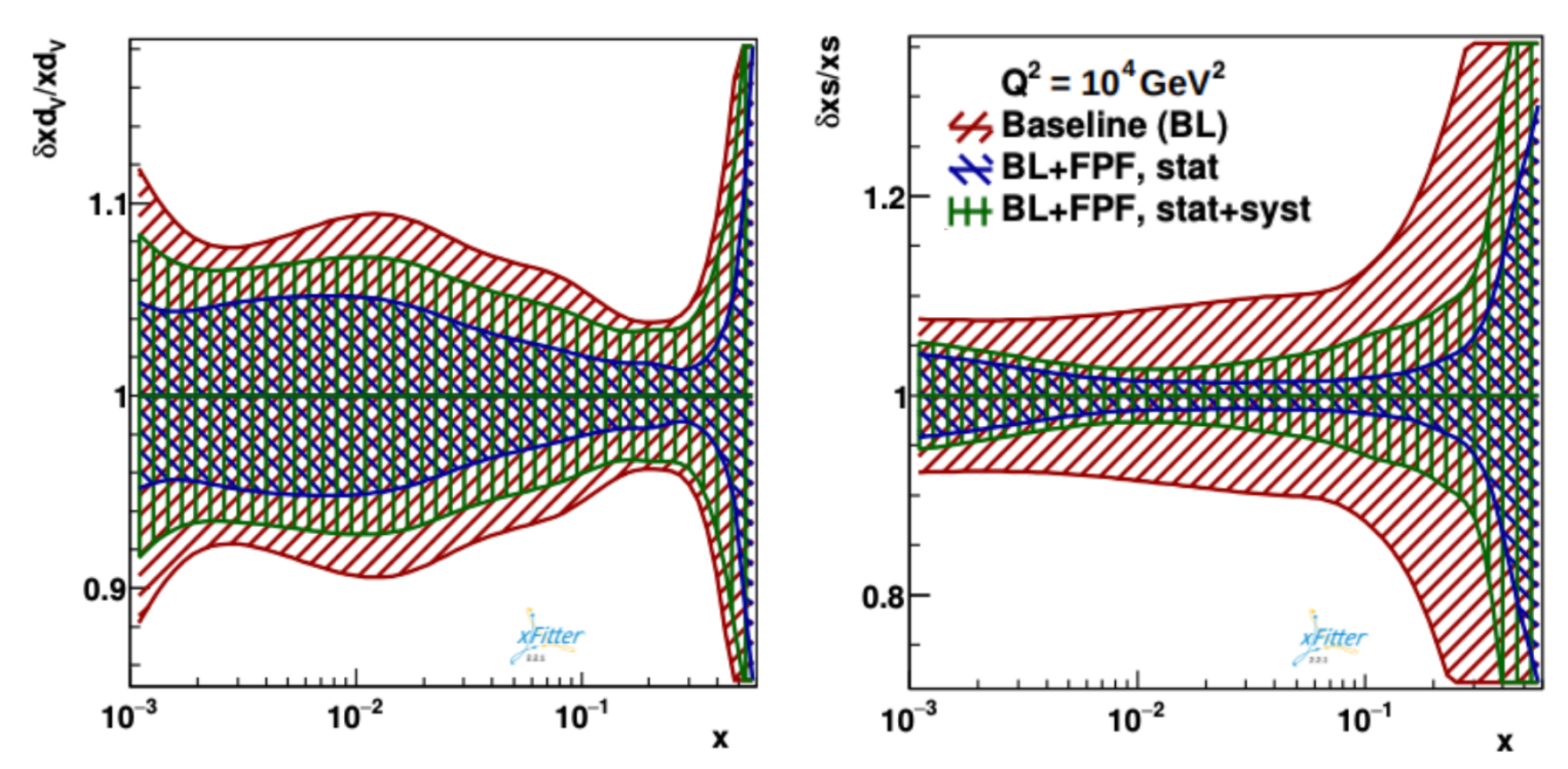}
    \caption{The impact of the combined FPF neutrino DIS dataset when added on top of the PDF4LHC21 baseline for the down valence $xd_V$ and the total strangeness $xs^+$ PDFs, for fits with and without including the projected systematic uncertainties.
    The bands correspond to the 68\% CL uncertainties.
    }
    \label{fig:fpf-impact-PDFs}
\end{figure}

By including these DIS projections in global (n)PDF analyses, specifically PDF4LHC21~\cite{PDF4LHCWorkingGroup:2022cjn}, NNPDF4.0~\cite{NNPDF:2021njg}, and EPPS21~\cite{Eskola:2021nhw}, it is found that neutrino structure function measurements at the LHC lead to significant reductions of both proton and nuclear PDF uncertainties.
This impact is illustrated in Fig.~\ref{fig:fpf-impact-PDFs}, showing the constraints from the combined FPF neutrino DIS dataset when added on top of the PDF4LHC21 baseline for the down valence $xd_V$ and the total strangeness $xs^+$ PDFs.
We consider the results of fits with and without including the projected systematic uncertainties.
This study also found that being able to tag $D$-mesons in the final state is necessary to achieve the ultimate reach on the strange PDF~\cite{Faura:2020oom}, and that neutrino charge flavour separation does not improve much the PDF sensitivity. 
In this context, further studies of neutrino DIS at the LHC will benefit from recent developments in event generators for high-energy neutrino scattering~\cite{Buonocore:2024pdv,FerrarioRavasio:2024kem}.

\begin{figure}[h]
    \centering
\includegraphics[width=0.65\textwidth,angle=-90]{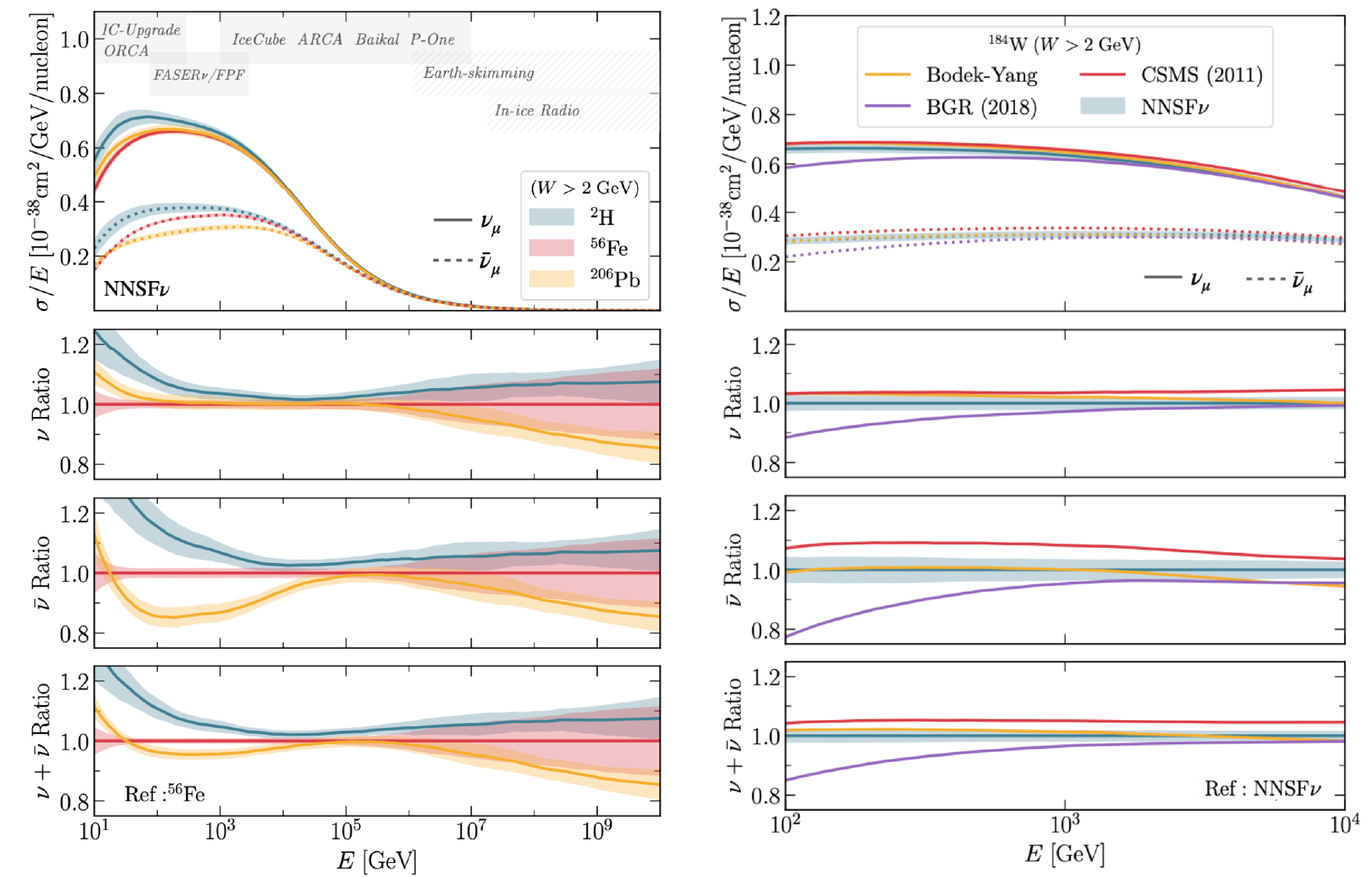}
    \caption{Left: the inclusive neutrino cross-section between $E_\nu=10$ GeV and $10^{10}$ GeV for different nuclear targets.
    The calculation is based on the NNSF$\nu$ framework and we indicate the relevant regions for different experiments. 
    Right: same now for a tungsten target and focusing on the energy region relevant for LHC far-forward experiments, comparing  NNSF$\nu$ with other calculations of the inclusive neutrino cross-section.
    }
    \label{fig:nnsfnu}
\end{figure}

The sensitivity of FPF measurements to different aspects of neutrino-nucleus scattering is also illustrated in  Fig.~\ref{fig:nnsfnu}.
The left panel shows the inclusive neutrino cross-section between $E_\nu=10$ GeV and $10^{10}$ GeV for different nuclear targets, separately for neutrinos and antineutrinos.
The calculation is based on the NNSF$\nu$ framework~\cite{Candido:2023utz} and we indicate the relevant regions for different experiments, from low to high energies. 
Nuclear effects are important for all $E_\nu$ regions, and hence should be accounted for when confronting with the experimental results.
The right panel focuses on the energy region relevant for LHC far-forward experiments and is provided on tungsten, comparing  NNSF$\nu$ with other calculations of the inclusive neutrino cross-section.
The observed differences between the various calculations would be resolved by the FPF measurements given the expected high neutrino yields~\cite{Feng:2022inv}.

\begin{figure}[h]
    \centering
\includegraphics[width=0.65\textwidth,angle=-90]{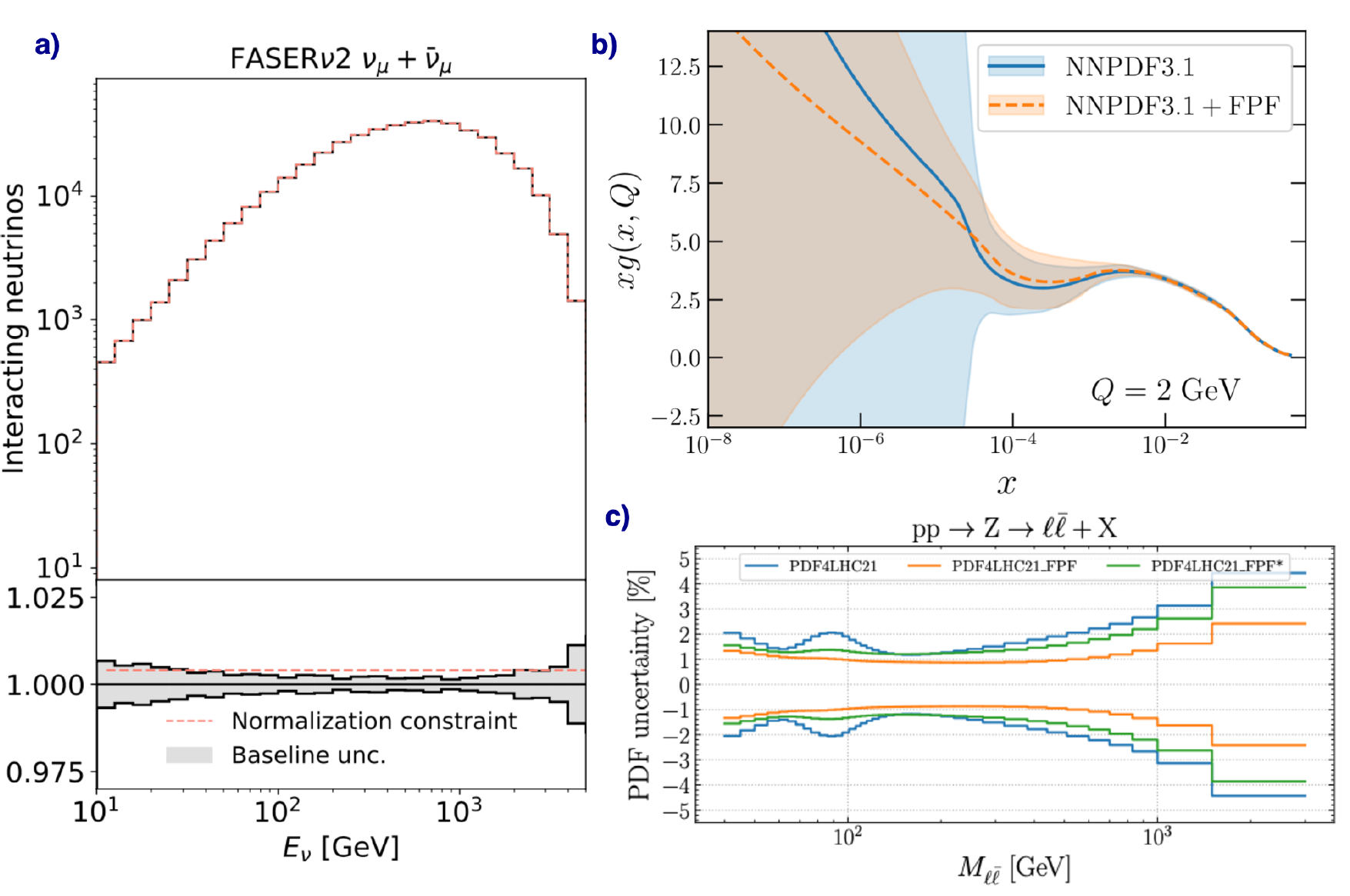}
    \caption{a) The normalization of the LHC forward neutrino flux can be determined simultaneously with DIS structure functions with permille precision at the FPF~\cite{Kling:2023tgr,Cruz-Martinez:2023sdv}. 
    b) Cross-section ratios between different neutrino rapidities enable constraints on the small-$x$ gluon PDF down to $x= 10^{-7}$.
    c) Proton PDFs informed by LHC neutrino data lead to improved predictions for high-$p_T$ backgrounds for BSM searches at the LHC. 
    }
    \label{fig:procs-fig3}
\end{figure}

\paragraph{Constraints on forward hadron production.}
Fig.~\ref{fig:procs-fig1} indicates how the LHC neutrino fluxes depend sensitively on the mechanisms for forward light- and heavy hadron-production in proton-proton collisions~\cite{Kling:2021gos,Buonocore:2023kna}. 
Theoretical predictions for forward hadron production at the LHC are affected by large uncertainties~\cite{FASER:2024ykc}, and hence {\it in situ} validation with experimental data becomes of utmost importance.
For instance, as demonstrated in~\cite{Kling:2023tgr,Cruz-Martinez:2023sdv}, measurements of the $E_\nu$ distribution at the LHC far-forward experiments can constrain the overall normalization of the $\nu_\mu+\bar{\nu}_\mu$ LHC forward neutrino flux at the permille level when fitted simultaneously with the DIS structure functions, see Fig.~\ref{fig:procs-fig3}a). 
It should be noted that disentangling physical effects in production and scattering is possible due to their different kinematic dependence, with production affecting mostly the $E_\nu$ and $y_\nu$ distributions while scattering modifying predominantly the $x$ and $Q^2$ distributions.

The significant theory uncertainties affecting forward $D$-meson production and fragmentation at the LHC difficult their use to constrain the small-$x$ gluon PDF.
In the same manner as for the case of $D$-meson production from LHCb~\cite{Gauld:2016kpd,Zenaiev:2019ktw}, to bypass this limitation one can construct dedicated observables where theory uncertainties partially cancel out without removing the PDF sensitivity. 
In the case of the LHC far-forward experiments, one such observable is provided by taking cross-section ratios between different rapidity regions,
\be
\mathcal{R}^{(\nu_e)}_{y_\nu}(E_\nu,\Delta_1 y_\nu,\Delta_2 y_\nu )\equiv N_{\nu_e}(E_\nu, \Delta_1 y_\nu)/N_{\nu_e}(E_\nu, \Delta_2 y_\nu) \, ,
\ee
with $\Delta_i y_\nu$ indicating a given neutrino rapidity region and  $N_{\nu_e}$ the number of interacting neutrinos within acceptance. 
To quantify the potential of observables such as $\mathcal{R}^{(\nu_e)}_{y_\nu}$ to pin down the small-$x$ PDFs, we generated projections for the FPF measurements of $\mathcal{R}^{(\nu_e)}_{y_\nu}$ (only statistical uncertainties were considered) and included them on the NNPDF3.1 set~\cite{NNPDF:2017mvq} by means of Bayesian reweighting, consistently with the procedure adopted in~\cite{Bertone:2018dse}.
The results of this study are shown in Fig.~\ref{fig:procs-fig3}b), confirming the potential of these optimised observables to access proton structure at very small-$x$ beyond the coverage of other laboratory experiments.
This region is particularly relevant for applications in astroparticle physics, for instance to constrain the prompt flux from charm decays~\cite{Gauld:2015yia} representing the main background for astrophysical neutrinos at IceCube and KM3NET.

\paragraph{Interplay with BSM searches at the HL-LHC.}
By shedding light on proton structure, neutrino DIS measurements at the LHC improve the prospects of BSM analyses at the HL-LHC.
First, because more precise PDFs lead to a reduction of theory uncertainties in processes such as Higgs production~\cite{LHCHiggsCrossSectionWorkingGroup:2016ypw}, a prime target of the HL-LHC in the search for anomalous Higgs couplings.
Second, FPF structure functions constrain large-$x$ PDFs, which in turn enable more robust predictions for high-$p_T$ backgrounds entering BSM searches at the LHC, such as the high-mass Drell-Yan distribution~\cite{Ball:2022qtp} shown in Fig.~\ref{fig:procs-fig3}c) and relevant for $W'$ and $Z'$ searches.

Third, neutrino measurements at the FPF provide a handle on the proton PDFs in the same large-$x$ region as high-$p_T$ measurements at the LHC but involving lower momentum transfers $Q^2$, hence breaking possible degeneracies between QCD and BSM effects which potentially affect the interpretation of high-$p_T$ data at the HL-LHC.
The interplay between PDFs and BSM in high-$p_T$ data has been quantified in a number of studies~\cite{Greljo:2021kvv,Kassabov:2023hbm,Hammou:2023heg,Gao:2022srd} using the SMEFT framework as a testbed.
These studies find that at the HL-LHC possible signals of BSM in Drell-Yan and top quark pair production could be reabsorbed into the PDFs without distorting the resultant fit quality.
With the motivation to prevent such scenario, a follow-up of the analysis of~\cite{Hammou:2023heg} demonstrates how including the FPF neutrino structure function projections of~\cite{Cruz-Martinez:2023sdv} within a global PDF fit breaks this PDF/BSM degeneracy and enables the robust identification of BSM signatures present in high-$p_T$ data~\cite{Hammou:2024}.

The analysis of~\cite{Hammou:2024} provides another independent motivation to carry out an extensive program of neutrino measurements at the HL-LHC: by enabling an independent calibration of the PDFs, the FPF ensures the optimal interpretation of high-$p_T$ measurements for BSM searches devoid of the possible contamination from PDF-related QCD effects.


\bibliographystyle{utphys}
\bibliography{DIS2024-nuDIS}

\providecommand{\href}[2]{#2}\begingroup\raggedright\begin{thebibliography}{10}

\bibitem{FASER:2023zcr}
{\bfseries FASER} Collaboration, H.~Abreu {\em et~al.}, ``{First Direct Observation of Collider Neutrinos with FASER at the LHC},'' \href{http://dx.doi.org/10.1103/PhysRevLett.131.031801}{{\em Phys. Rev. Lett.} {\bfseries 131} no.~3, (2023) 031801}, \href{http://arxiv.org/abs/2303.14185}{{\ttfamily arXiv:2303.14185 [hep-ex]}}.

\bibitem{SNDLHC:2023pun}
{\bfseries SND@LHC} Collaboration, R.~Albanese {\em et~al.}, ``{Observation of Collider Muon Neutrinos with the SND@LHC Experiment},'' \href{http://dx.doi.org/10.1103/PhysRevLett.131.031802}{{\em Phys. Rev. Lett.} {\bfseries 131} no.~3, (2023) 031802}, \href{http://arxiv.org/abs/2305.09383}{{\ttfamily arXiv:2305.09383 [hep-ex]}}.

\bibitem{FASER:2024hoe}
{\bfseries FASER} Collaboration, R.~Mammen~Abraham {\em et~al.}, ``{First Measurement of the $\nu_e$ and $\nu_\mu$ Interaction Cross Sections at the LHC with FASER's Emulsion Detector},'' \href{http://arxiv.org/abs/2403.12520}{{\ttfamily arXiv:2403.12520 [hep-ex]}}.

\bibitem{Anchordoqui:2021ghd}
L.~A. Anchordoqui {\em et~al.}, ``{The Forward Physics Facility: Sites, experiments, and physics potential},'' \href{http://dx.doi.org/10.1016/j.physrep.2022.04.004}{{\em Phys. Rept.} {\bfseries 968} (2022) 1--50}, \href{http://arxiv.org/abs/2109.10905}{{\ttfamily arXiv:2109.10905 [hep-ph]}}.

\bibitem{Feng:2022inv}
J.~L. Feng {\em et~al.}, ``{The Forward Physics Facility at the High-Luminosity LHC},'' \href{http://dx.doi.org/10.1088/1361-6471/ac865e}{{\em J. Phys. G} {\bfseries 50} no.~3, (2023) 030501}, \href{http://arxiv.org/abs/2203.05090}{{\ttfamily arXiv:2203.05090 [hep-ex]}}.

\bibitem{Batell:2021blf}
B.~Batell, J.~L. Feng, and S.~Trojanowski, ``{Detecting Dark Matter with Far-Forward Emulsion and Liquid Argon Detectors at the LHC},'' \href{http://dx.doi.org/10.1103/PhysRevD.103.075023}{{\em Phys. Rev. D} {\bfseries 103} no.~7, (2021) 075023}, \href{http://arxiv.org/abs/2101.10338}{{\ttfamily arXiv:2101.10338 [hep-ph]}}.

\bibitem{Kling:2021gos}
F.~Kling and L.~J. Nevay, ``{Forward neutrino fluxes at the LHC},'' \href{http://dx.doi.org/10.1103/PhysRevD.104.113008}{{\em Phys. Rev. D} {\bfseries 104} no.~11, (2021) 113008}, \href{http://arxiv.org/abs/2105.08270}{{\ttfamily arXiv:2105.08270 [hep-ph]}}.

\bibitem{Buonocore:2023kna}
L.~Buonocore, F.~Kling, L.~Rottoli, and J.~Sominka, ``{Predictions for neutrinos and new physics from forward heavy hadron production at the LHC},'' \href{http://dx.doi.org/10.1140/epjc/s10052-024-12726-5}{{\em Eur. Phys. J. C} {\bfseries 84} no.~4, (2024) 363}, \href{http://arxiv.org/abs/2309.12793}{{\ttfamily arXiv:2309.12793 [hep-ph]}}.

\bibitem{IceCube:2017lak}
{\bfseries IceCube} Collaboration, M.~G. Aartsen {\em et~al.}, ``{Measurement of Atmospheric Neutrino Oscillations at 6\textendash{}56 GeV with IceCube DeepCore},'' \href{http://dx.doi.org/10.1103/PhysRevLett.120.071801}{{\em Phys. Rev. Lett.} {\bfseries 120} no.~7, (2018) 071801}, \href{http://arxiv.org/abs/1707.07081}{{\ttfamily arXiv:1707.07081 [hep-ex]}}.

\bibitem{IceCube:2014stg}
{\bfseries IceCube} Collaboration, M.~G. Aartsen {\em et~al.}, ``{Observation of High-Energy Astrophysical Neutrinos in Three Years of IceCube Data},'' \href{http://dx.doi.org/10.1103/PhysRevLett.113.101101}{{\em Phys. Rev. Lett.} {\bfseries 113} (2014) 101101}, \href{http://arxiv.org/abs/1405.5303}{{\ttfamily arXiv:1405.5303 [astro-ph.HE]}}.

\bibitem{KM3NeT:2021ozk}
{\bfseries KM3NeT} Collaboration, S.~Aiello {\em et~al.}, ``{Determining the neutrino mass ordering and oscillation parameters with KM3NeT/ORCA},'' \href{http://dx.doi.org/10.1140/epjc/s10052-021-09893-0}{{\em Eur. Phys. J. C} {\bfseries 82} no.~1, (2022) 26}, \href{http://arxiv.org/abs/2103.09885}{{\ttfamily arXiv:2103.09885 [hep-ex]}}.

\bibitem{KM3Net:2016zxf}
{\bfseries KM3Net} Collaboration, S.~Adrian-Martinez {\em et~al.}, ``{Letter of intent for KM3NeT 2.0}'' \href{http://dx.doi.org/10.1088/0954-3899/43/8/084001}{{\em J. Phys. G} {\bfseries 43} no.~8, (2016) 084001}, \href{http://arxiv.org/abs/1601.07459}{{\ttfamily arXiv:1601.07459 [astro-ph.IM]}}.

\bibitem{Cruz-Martinez:2023sdv}
J.~M. Cruz-Martinez, M.~Fieg, T.~Giani, P.~Krack, T.~M\"akel\"a, T.~R. Rabemananjara, and J.~Rojo, ``{The LHC as a Neutrino-Ion Collider},'' \href{http://dx.doi.org/10.1140/epjc/s10052-024-12665-1}{{\em Eur. Phys. J. C} {\bfseries 84} no.~4, (2024) 369}, \href{http://arxiv.org/abs/2309.09581}{{\ttfamily arXiv:2309.09581 [hep-ph]}}.

\bibitem{AbdulKhalek:2018rok}
R.~Abdul~Khalek, S.~Bailey, J.~Gao, L.~Harland-Lang, and J.~Rojo, ``{Towards Ultimate Parton Distributions at the High-Luminosity LHC},'' \href{http://dx.doi.org/10.1140/epjc/s10052-018-6448-y}{{\em Eur. Phys. J. C} {\bfseries 78} no.~11, (2018) 962}, \href{http://arxiv.org/abs/1810.03639}{{\ttfamily arXiv:1810.03639 [hep-ph]}}.

\bibitem{AbdulKhalek:2021gbh}
R.~Abdul~Khalek {\em et~al.}, ``{Science Requirements and Detector Concepts for the Electron-Ion Collider}: {EIC Yellow Report},'' \href{http://dx.doi.org/10.1016/j.nuclphysa.2022.122447}{{\em Nucl. Phys. A} {\bfseries 1026} (2022) 122447}, \href{http://arxiv.org/abs/2103.05419}{{\ttfamily arXiv:2103.05419 [physics.ins-det]}}.

\bibitem{PDF4LHCWorkingGroup:2022cjn}
{\bfseries PDF4LHC Working Group} Collaboration, R.~D. Ball {\em et~al.}, ``{The PDF4LHC21 combination of global PDF fits for the LHC Run III},'' \href{http://dx.doi.org/10.1088/1361-6471/ac7216}{{\em J. Phys. G} {\bfseries 49} no.~8, (2022) 080501}, \href{http://arxiv.org/abs/2203.05506}{{\ttfamily arXiv:2203.05506 [hep-ph]}}.

\bibitem{NNPDF:2021njg}
{\bfseries NNPDF} Collaboration, R.~D. Ball {\em et~al.}, ``{The path to proton structure at 1\% accuracy},'' \href{http://dx.doi.org/10.1140/epjc/s10052-022-10328-7}{{\em Eur. Phys. J. C} {\bfseries 82} no.~5, (2022) 428}, \href{http://arxiv.org/abs/2109.02653}{{\ttfamily arXiv:2109.02653 [hep-ph]}}.

\bibitem{Eskola:2021nhw}
K.~J. Eskola, P.~Paakkinen, H.~Paukkunen, and C.~A. Salgado, ``{EPPS21: a global QCD analysis of nuclear PDFs},'' \href{http://dx.doi.org/10.1140/epjc/s10052-022-10359-0}{{\em Eur. Phys. J. C} {\bfseries 82} no.~5, (2022) 413}, \href{http://arxiv.org/abs/2112.12462}{{\ttfamily arXiv:2112.12462 [hep-ph]}}.

\bibitem{Faura:2020oom}
F.~Faura, S.~Iranipour, E.~R. Nocera, J.~Rojo, and M.~Ubiali, ``{The Strangest Proton?},'' \href{http://dx.doi.org/10.1140/epjc/s10052-020-08749-3}{{\em Eur. Phys. J. C} {\bfseries 80} no.~12, (2020) 1168}, \href{http://arxiv.org/abs/2009.00014}{{\ttfamily arXiv:2009.00014 [hep-ph]}}.

\bibitem{Buonocore:2024pdv}
L.~Buonocore, G.~Limatola, P.~Nason, and F.~Tramontano, ``{An event generator for Lepton-Hadron Deep Inelastic Scattering at NLO+PS with POWHEG including mass effects},'' \href{http://arxiv.org/abs/2406.05115}{{\ttfamily arXiv:2406.05115 [hep-ph]}}.

\bibitem{FerrarioRavasio:2024kem}
S.~Ferrario~Ravasio, R.~Gauld, B.~J\"ager, A.~Karlberg, and G.~Zanderighi, ``{An event generator for neutrino-induced Deep Inelastic Scattering and applications to neutrino astronomy},'' \href{http://arxiv.org/abs/2407.03894}{{\ttfamily arXiv:2407.03894 [hep-ph]}}.

\bibitem{Candido:2023utz}
A.~Candido, A.~Garcia, G.~Magni, T.~Rabemananjara, J.~Rojo, and R.~Stegeman, ``{Neutrino Structure Functions from GeV to EeV Energies},'' \href{http://dx.doi.org/10.1007/JHEP05(2023)149}{{\em JHEP} {\bfseries 05} (2023) 149}, \href{http://arxiv.org/abs/2302.08527}{{\ttfamily arXiv:2302.08527 [hep-ph]}}.

\bibitem{FASER:2024ykc}
{\bfseries FASER} Collaboration, R.~Mammen~Abraham {\em et~al.}, ``{Neutrino Rate Predictions for FASER},'' \href{http://arxiv.org/abs/2402.13318}{{\ttfamily arXiv:2402.13318 [hep-ex]}}.

\bibitem{Kling:2023tgr}
F.~Kling, T.~M\"akel\"a, and S.~Trojanowski, ``{Investigating the fluxes and physics potential of LHC neutrino experiments},'' \href{http://dx.doi.org/10.1103/PhysRevD.108.095020}{{\em Phys. Rev. D} {\bfseries 108} no.~9, (2023) 095020}, \href{http://arxiv.org/abs/2309.10417}{{\ttfamily arXiv:2309.10417 [hep-ph]}}.

\bibitem{Gauld:2016kpd}
R.~Gauld and J.~Rojo, ``{Precision determination of the small-$x$ gluon from charm production at LHCb},'' \href{http://dx.doi.org/10.1103/PhysRevLett.118.072001}{{\em Phys. Rev. Lett.} {\bfseries 118} no.~7, (2017) 072001}, \href{http://arxiv.org/abs/1610.09373}{{\ttfamily arXiv:1610.09373 [hep-ph]}}.

\bibitem{Zenaiev:2019ktw}
{\bfseries PROSA} Collaboration, O.~Zenaiev, M.~V. Garzelli, K.~Lipka, S.~O. Moch, A.~Cooper-Sarkar, F.~Olness, A.~Geiser, and G.~Sigl, ``{Improved constraints on parton distributions using LHCb, ALICE and HERA heavy-flavour measurements and implications for the predictions for prompt atmospheric-neutrino fluxes},'' \href{http://dx.doi.org/10.1007/JHEP04(2020)118}{{\em JHEP} {\bfseries 04} (2020) 118}, \href{http://arxiv.org/abs/1911.13164}{{\ttfamily arXiv:1911.13164 [hep-ph]}}.

\bibitem{NNPDF:2017mvq}
{\bfseries NNPDF} Collaboration, R.~D. Ball {\em et~al.}, ``{Parton distributions from high-precision collider data},'' \href{http://dx.doi.org/10.1140/epjc/s10052-017-5199-5}{{\em Eur. Phys. J. C} {\bfseries 77} no.~10, (2017) 663}, \href{http://arxiv.org/abs/1706.00428}{{\ttfamily arXiv:1706.00428 [hep-ph]}}.

\bibitem{Bertone:2018dse}
V.~Bertone, R.~Gauld, and J.~Rojo, ``{Neutrino Telescopes as QCD Microscopes},'' \href{http://dx.doi.org/10.1007/JHEP01(2019)217}{{\em JHEP} {\bfseries 01} (2019) 217}, \href{http://arxiv.org/abs/1808.02034}{{\ttfamily arXiv:1808.02034 [hep-ph]}}.

\bibitem{Gauld:2015yia}
R.~Gauld, J.~Rojo, L.~Rottoli, and J.~Talbert, ``{Charm production in the forward region: constraints on the small-x gluon and backgrounds for neutrino astronomy},'' \href{http://dx.doi.org/10.1007/JHEP11(2015)009}{{\em JHEP} {\bfseries 11} (2015) 009}, \href{http://arxiv.org/abs/1506.08025}{{\ttfamily arXiv:1506.08025 [hep-ph]}}.

\bibitem{LHCHiggsCrossSectionWorkingGroup:2016ypw}
{\bfseries LHC Higgs Cross Section Working Group} Collaboration, D.~de~Florian {\em et~al.}, ``{Handbook of LHC Higgs Cross Sections: 4. Deciphering the Nature of the Higgs Sector},'' \href{http://arxiv.org/abs/1610.07922}{{\ttfamily arXiv:1610.07922 [hep-ph]}}.

\bibitem{Ball:2022qtp}
R.~D. Ball, A.~Candido, S.~Forte, F.~Hekhorn, E.~R. Nocera, J.~Rojo, and C.~Schwan, ``{Parton distributions and new physics searches: the Drell\textendash{}Yan forward\textendash{}backward asymmetry as a case study},'' \href{http://dx.doi.org/10.1140/epjc/s10052-022-11133-y}{{\em Eur. Phys. J. C} {\bfseries 82} no.~12, (2022) 1160}, \href{http://arxiv.org/abs/2209.08115}{{\ttfamily arXiv:2209.08115 [hep-ph]}}.

\bibitem{Greljo:2021kvv}
A.~Greljo, S.~Iranipour, Z.~Kassabov, M.~Madigan, J.~Moore, J.~Rojo, M.~Ubiali, and C.~Voisey, ``{Parton distributions in the SMEFT from high-energy Drell-Yan tails},'' \href{http://dx.doi.org/10.1007/JHEP07(2021)122}{{\em JHEP} {\bfseries 07} (2021) 122}, \href{http://arxiv.org/abs/2104.02723}{{\ttfamily arXiv:2104.02723 [hep-ph]}}.

\bibitem{Kassabov:2023hbm}
Z.~Kassabov, M.~Madigan, L.~Mantani, J.~Moore, M.~M. Alvarado, J.~Rojo, and M.~Ubiali, ``{The top quark legacy of the LHC Run II for PDF and SMEFT analyses},'' \href{http://arxiv.org/abs/2303.06159}{{\ttfamily arXiv:2303.06159 [hep-ph]}}.

\bibitem{Hammou:2023heg}
E.~Hammou, Z.~Kassabov, M.~Madigan, M.~L. Mangano, L.~Mantani, J.~Moore, M.~M. Alvarado, and M.~Ubiali, ``{Hide and seek: how PDFs can conceal new physics},'' \href{http://dx.doi.org/10.1007/JHEP11(2023)090}{{\em JHEP} {\bfseries 11} (2023) 090}, \href{http://arxiv.org/abs/2307.10370}{{\ttfamily arXiv:2307.10370 [hep-ph]}}.

\bibitem{Gao:2022srd}
J.~Gao, M.~Gao, T.~J. Hobbs, D.~Liu, and X.~Shen, ``{Simultaneous CTEQ-TEA extraction of PDFs and SMEFT parameters from jet and $ t\overline{t} $ data},'' \href{http://dx.doi.org/10.1007/JHEP05(2023)003}{{\em JHEP} {\bfseries 05} (2023) 003}, \href{http://arxiv.org/abs/2211.01094}{{\ttfamily arXiv:2211.01094 [hep-ph]}}.

\bibitem{Hammou:2024}
E.~Hammou, Madigan, and M.~Ubiali, ``{in preparation},''.

\end{thebibliography}\endgroup

\end{document}